\documentclass[onecolumn]{aastex63}
\usepackage[utf8]{inputenc}
\usepackage{amssymb}
\usepackage{gensymb}
\usepackage{amsmath}
\usepackage{upgreek}
\usepackage{mathtools}
\usepackage{array}
\usepackage{xcolor}
\usepackage{xfrac}
\usepackage{natbib}
\usepackage{listings}
\usepackage[none]{hyphenat}
\usepackage{hyperref}
\usepackage{tikz}
\usepackage{cancel}
\usetikzlibrary{automata,positioning,decorations.markings,calc,arrows,angles,quotes}

\newcommand\tab[1][0.5cm]{\hspace*{#1}}

\definecolor{darkblue}{RGB}{0, 39, 102}
\definecolor{section}{RGB}{0, 0, 0}
\definecolor{subsection}{RGB}{0, 0, 75}
\definecolor{subsubsection}{RGB}{0, 75, 0}
\definecolor{lightblue}{RGB}{94, 215, 255}
\definecolor{onefinite}{RGB}{173, 99, 19}
\definecolor{twofinite}{RGB}{0, 125, 50}
\definecolor{lightgray}{RGB}{150, 150, 150}
\definecolor{lightcyan}{RGB}{105, 193, 201}

\hypersetup{hidelinks, colorlinks=true, linkcolor=blue, filecolor=magenta, urlcolor=blue, citecolor=blue}

\newcommand{\mbeq}{\overset{!}{=}}

\newcommand{\minus}{\raisebox{.2\height}{\scalebox{.7}{$-$}}}
\newcommand{\pcos}[1]{\cos \left( #1 \right)}
\newcommand{\psin}[1]{\sin \left( #1 \right)}
\newcommand{\ptan}[1]{\tan \left( #1 \right)}
\newcommand{\atan}[1]{\tan^{\minus 1} \left( #1 \right)}
\newcommand{\acot}[1]{\cot^{\minus 1} \left( #1 \right)}
\newcommand{\atantwo}[1]{\text{atan2} \left( #1 \right)}
\newcommand{\acos}[1]{\cos^{\minus 1} \left( #1 \right)}
\newcommand{\asin}[1]{\sin^{\minus 1} \left( #1 \right)}

\newcommand{\sci}[2]{#1 \!\times\! 10^{#2}}

\newcommand{\changed}[1]{#1}

\newcommand{\spec}{\text{spec}}
\newcommand{\src}{\text{src}}
\newcommand{\obs}{\text{obs}}
\newcommand{\sph}{\text{sph}}

\newcommand{\onefinite}{{\color{onefinite} \textbf{one-finite}}}

\newcommand{\twofinite}{{\color{twofinite} \textbf{both-finite}}}

\newenvironment{nscenter}
 {\setlength{\topsep}{0cm}\trivlist\item\relax\centering}
 {\endtrivlist}

\def\carc[#1] (#2)(#3:#4:#5) {
	\draw[#1] ($(#2)+({#5*cos(#3)}, {#5*sin(#3)})$) arc (#3:#4:#5);
}

\shorttitle{Solving the Alhazen-Ptolemy Problem}
\shortauthors{Miller, Barnes, and MacKenzie}

\begin{document}
\title{Solving the Alhazen-Ptolemy Problem: Determining Specular Points 
       on Spherical Surfaces for Radiative Transfer of Titan's Seas}
\author[0000-0002-1525-0972]{William J. Miller}
\affiliation{University of Idaho, Department of Physics}

\author[0000-0002-7755-3530]{Jason W. Barnes}
\affiliation{University of Idaho, Department of Physics}

\author[0000-0002-1658-9687]{Shannon M. MacKenzie}
\affiliation{Johns Hopkins University, Applied Physics Laboratory}

\keywords{specular reflections}

\begin{abstract}
Given a light source, a spherical reflector, and an observer, where on the surface 
of the sphere will the light be directly reflected to the observer, i.e. where is 
the the specular point? This is known as the Alhazen-Ptolemy problem, and finding 
this specular point for spherical reflectors is useful in applications ranging 
from computer rendering to atmospheric modeling to GPS communications. Existing 
solutions rely upon finding the roots of a quartic equation and evaluating 
numerically which root provides the real specular point. We offer a formulation, 
and two solutions thereof, for which the correct root is predeterminable, thereby 
allowing the construction of the fully analytical solutions we present. Being 
faster to compute, our solutions should prove useful in cases which require 
repeated calculation of the specular point, such as Monte-Carlo radiative 
transfer, including reflections off of Titan's hydrocarbon seas.
\end{abstract}

\begin{figure}[b]
	\centering
	\begin{minipage}[b][][b]{0.45\textwidth}
		\begin{tikzpicture}
			\coordinate (orig) at (0,0);
			\coordinate (l) at (-4, 0);
			\coordinate (r) at (4, 0);
			\coordinate (v) at (1, 3);
			\coordinate (sp) at (1, 0);
			\node (s) at (-4, 3) {$\bullet$};
			\node (d) at (4, 2) {$\bullet$};
			\node (sy) at (-4, 1.5) {$y_1$};
			\node (sx) at (-1.5, -0.25) {$x_1$};
			\node (dy) at (4, 1) {$y_2$};
			\node (dx) at (2.5, -0.25) {$x_2$};
			\node (xs) at (0, 3.25) {$x_s$};
			
			\draw [line width = 0.75pt] (l) to (r);
% 			\draw [line width = 1pt, dotted] (sp) to (v);
			\draw [line width = 1pt, arrows=-{latex'}] (s) to (sp);
			\draw [line width = 1pt, arrows=-{latex'}] (sp) to (d);
			\draw [line width = 1pt, arrows=-{latex'}, dotted] (xs) to +(-4, 0);
			\draw [line width = 1pt, arrows=-{latex'}, dotted] (xs) to +(4, 0);
			
			\draw [line width = 0.75pt, arrows=-{latex'}, dotted] (sy) to (l);
			\draw [line width = 0.75pt, arrows=-{latex'}, dotted] (sy) to (s);
			\draw [line width = 0.75pt, arrows=-{latex'}, dotted] (sx) -- +(-2.5,0);
			\draw [line width = 0.75pt, arrows=-{latex'}, dotted] (sx) -- +(2.5,0);
			
			\draw [line width = 0.75pt, arrows=-{latex'}, dotted] (dy) to (r);
			\draw [line width = 0.75pt, arrows=-{latex'}, dotted] (dy) to (d);
			\draw [line width = 0.75pt, arrows=-{latex'}, dotted] (dx) -- +(-1.5,0);
			\draw [line width = 0.75pt, arrows=-{latex'}, dotted] (dx) -- +(1.5,0);
			
			\pic[draw, -, "$\theta_e$", angle eccentricity = 1.2, angle radius = 1cm] {angle = r--sp--d};
			\pic[draw, -, "$\theta_i$", angle eccentricity = 1.2, angle radius = 1cm] {angle = s--sp--l};
			
		\end{tikzpicture}
		\caption{Specular point for planar surface.}
% 		 $(x_1, y_1)$ is the position of the source
% 	     and $(x_2, y_2)$ is the position of the observer in the coordinate system 
% 	     centered on the specular point.
		\label{fig:PlanarSpecular}
	\end{minipage} \hfill
	\begin{minipage}[b][][b]{0.5\textwidth}
		\resizebox{\textwidth}{!} {
			\begin{tikzpicture}
				\node[inner sep=0pt] (orig) at (0,0) {$\bullet$};
				\coordinate (right) at (5, 0);
				\node (titan) at (0, -0.5) {};
				\node[inner sep=0pt] (scatter) at (0, 8) {$\bullet$};
				\node (scatterlabel) at (0, 8.5) {Source};
				\node[inner sep=0pt] (specular) at (2.0404, 4.5647) {$\bullet$};
				
				\coordinate (vspecular) at (2.0404, 8);
				\coordinate (hspecular) at (6, 4.5647);
				\coordinate (lspecular) at (1.0804, 4.90672);
				\coordinate (rspecular) at (3.0404, 4.10267);
				
				\node[label={[align=center] Specular \\ Point}] (specularlabel) at (1., 3.5) {};
				\node[label={[align=center] Observer}] (detectorlabel) at (8.5, 9.25) {};
				\node (rd) at (5.25, 4) {$R_{\obs}$};
				\node (rt) at (-0.5, 2.5) {$R_{\sph}$};
				\node (rs) at (-1, 4) {$R_{\src}$};
				\node (detector) at (10, 9.5647) {$\bullet$};
				\node (detector2) at (10, 9.5647) {};
				
				\carc[line width = 0.75pt] (orig)(-5:105:5)
				
				\draw[line width = 0.75pt, dotted] (orig) to (detector);
				\draw[line width = 0.75pt, dotted] (orig) to (scatter);
				\draw[line width = 0.75pt, dotted] (orig) to (specular);
				\draw[line width = 0.75pt, dotted] (orig) to (right);
% 				\draw[line width = 0.75pt, dotted] (specular) to (vspecular);
				\draw[line width = 0.75pt, dotted] (specular) to (hspecular);
				\draw[line width = 0.75pt, arrows = -{latex'}] (scatter) to (specular);
				\draw[line width = 0.75pt, arrows = -{latex'}] (specular) to (detector2);
				\draw[line width = 1pt] (0, 0.25) -- +(0.25, 0);
				\draw[line width = 1pt] (0.25, 0) -- +(0, 0.25);
				\draw[line width = 0.75pt, arrows = -{latex'}] (rs) -- +(0, 4);
				\draw[line width = 0.75pt, arrows = -{latex'}] (rs) -- +(0, -4);
				\draw[line width = 0.75pt, arrows = -{latex'}] (rt) -- +(0, 2.5);
				\draw[line width = 0.75pt, arrows = -{latex'}] (rt) -- +(0, -2.5);
				
				\pic[draw, -, "$\theta_{\obs}$", angle eccentricity = 1.35, angle radius = 1cm] {angle = right--orig--detector};
				\pic[draw, -, "$\theta_{\spec}$", angle eccentricity = 1.15, angle radius = 2cm] {angle = detector--orig--specular};
				\pic[draw, -, "", angle eccentricity = 1.5, angle radius = 2cm] {angle = right--orig--detector};
				\pic[draw, -, "$\theta_e$", angle eccentricity = 1.5, angle radius = 0.6cm] {angle = rspecular--specular--hspecular};
				\pic[draw, -, "", angle eccentricity = 1.5, angle radius = 0.6cm] {angle = hspecular--specular--detector2};
				\pic[draw, -, "$\theta_i$", angle eccentricity = 1.5, angle radius = 0.75cm] {angle = scatter--specular--lspecular};
				\pic[draw, -, "$\theta_2$", angle eccentricity = 1.35, angle radius = 1.cm] {angle = detector2--specular--vspecular};
				\pic[draw, -, "", angle eccentricity = 1.35, angle radius = 1.cm] {angle = hspecular--specular--scatter};
				\pic[draw, -, "$\theta_1$", angle eccentricity = 1.2, angle radius = 1.5cm] {angle = hspecular--specular--detector2};
			\end{tikzpicture}
		}
		\caption{Specular point on a spherical surface.} 
		\label{fig:SphericalSpecular}
	\end{minipage}
\end{figure}

\section{Introduction} \label{sec:intro}

The specular point is the specific location on a surface where the outgoing vector of  
a mirror reflection coincides exactly with the vector to an observer. At such a point, 
the light \changed{from a point source is seen by the observer} and, 
if the source has spatial extent, an image is produced in the reflection. Since 
superposition allows a source with spatial extent to be divided into point sources -- 
each with a single specular point which in aggregate comprise the specular region -- 
the same methods for finding a specular point can be repeatedly applied to determine 
the region of specular reflection corresponding to a source with spatial extent.

\subsection{Applications}
Perhaps the most notable use for finding the specular point for an arbitrary 
source-observer configuration is in Global Positioning Systems. Unfortunately, the 
tolerance of GPS requires the Earth be approximated as an ellipsoid instead of a 
sphere, rendering analytical solution of the spherical-surface case less useful 
\citep{southwell2018new, prakash1994algorithm}. However, our formulation can be 
extended to elliptical geometries more easily than previous approaches, arguing 
positively of its usefulness in this application. 

Unlike the Earth, Saturn's moon Titan -- which harbors liquid hydrocarbons that 
allowed the Cassini mission to record numerous specular reflections
\citep{stephan2010specular, barnes2014cassini} -- can be well-approximated as a 
sphere. The \lstinline{SRTC++}{} model \citep{barnes2018spherical} does exactly 
that and finding a faster method for computing the specular reflections due to 
scattering off of Titan's ubiquitous atmospheric haze aerosols in 
\lstinline{SRTC++}{} is the direct motivation for this work. There may also be 
utility for removing `sun-glint' from imaging data by employing a method similar 
to those in \citet{kay2009sun} but with the higher precision enabled by our 
analytical solutions.

Another notable application is the rendering of computer graphics with raytracing 
\citep{inakage1986caustics}, a case where computation speed is critical. 
Particularly when rendering a reflected image (instead of a simple point 
source) because the rendering fidelity is much more sensitive to the accuracy of 
the specular point calculation. The increased interest in ray-traced rendering,
particularly with regard to the video game and movie industries, serves only to 
increase the usefulness of a faster method for handling even partially reflective 
surfaces. 

\subsection{Planar Surface}

For a planar surface the specular point (Figure \ref{fig:PlanarSpecular}) can be 
found trivially. In the coordinate system centered on the specular point with the 
$x$-axis parallel to the surface, the positions of the source and observer must
satisfy 
\begin{equation} \label{eqn:IE}
	\theta_i \mbeq \theta_e
\end{equation}
where $\theta_i$ is incidence angle and $\theta_e$ is emission angle. In the 
configuration shown in Figure \ref{fig:PlanarSpecular} they are defined as
\begin{subequations}
	\begin{equation*}
		\theta_i = \atan{\frac{y_1}{x_1}}, \tab \text{for} \tab 0 \leq \theta_i \leq \frac{\pi}{2},
	\end{equation*}
	\begin{equation*}
		\theta_e = \atan{\frac{y_2}{x_2}}, \tab \text{for} \tab 0 \leq \theta_e \leq \frac{\pi}{2}, \tab \text{and}
	\end{equation*}
	\begin{equation*}
		x_s = x_1 + x_2.
	\end{equation*}
\end{subequations}
where $x_s$ is the separation between the source and observer in the $x$-direction. 
Then the position of the source and observer 
relative to the specular point can be found by combining the above and rearranging,
\begin{subequations}
	\begin{equation*}
		x_2 = \frac{x_s y_2}{y_2 - y_1}, 
	\end{equation*}
	\begin{equation*}
		x_1 = \frac{x_s y_1}{y_1 - y_2}.
	\end{equation*}
\end{subequations}

\subsection{Alhazen-Ptolemy Problem}

The case of a spherical surface (Figure \ref{fig:SphericalSpecular}) is far more 
complicated. First formulated in 150 CE by Ptolemy \citep{weisstein2002alhazen}, 
it is known as the Alhazen-Ptolemy problem because the first method of solving it 
(a geometric approach using conic sections) was provided -- and proven -- by 
\textit{Ibn al-Haythum} in the 11$^{th}$ century \citep{katz1995ideas}. 
\cite{elkin1965deceptively} provided the first algebraic solution and similar 
formulations by \citet{riede1989reflexion}, \citet{smith1992remarkable}, and 
\citet{waldvogel1992problem} confirmed Elkin's solution. \cite{neumann1998reflections} 
proved that ruler-and-compass construction of a general solution, a method which 
permits only circular arcs and straight lines, is impossible. That is, given only 
an unmarked and idealized compass and straight edge there is no way to determine 
the specular point. 

\subsubsection{Existing methods}

The solution in \citet{elkin1965deceptively} produces a quartic equation for 
the distance from either the source or the observer to the specular point 
(though the he does not term it such), the roots of which can be used to retrieve 
a collection of distances from the specular point to the other point of interest. 
The resulting $x$-$y$ pairs must then be analyzed to find the real specular 
point. There are several methods which are useful in retrieving the specular point
from the generated $x$-$y$ pairs: minimizing the source-specular-observer 
path length, employing a numerical root finding method with an initial `guess'
(e.g. Newton's method), or `by probe'.

\citet{fujimura2019ptolemy} provides a formulation which works well for the 
method of path-length minimization, but suffers numerical instability in cases
where the radius of the sphere is very small relative to source or observer distance
(e.g. in the {\onefinite} case discussed below). \cite{glaeser1999reflections} 
emonstrates the numerical root finding approach (using an algorithm from 
\citealt{schwarze1990cubic}) and notes that it is subject to ``numerical 
instabilities that may lead to a considerable loss of accuracy" under certain 
circumstances. \citet{elkin1965deceptively} shows the `by probe' method, but 
this is almost impossible to capture algorithmically. These limitations beg the
creation of a better, faster method.

\subsubsection{Our approach}

The salient problem, and in fact the only obstacle to a fully analytical solution, 
is that of deducing which root to use. We present a formulation that allows 
predetermination of the `correct' root and enables the construction of a piecewise 
function which directly produces the location of the physical specular point.
We find this formulation superior to previous work because of its analytical 
branch deduction -- something lacking from every other formulation. The fully
analytical approach avoids numerical instability and allows significantly faster 
computation.

We present two analytical solutions for this formulation: one that requires 
the distance to either the source or observer be approximateable as infinite (the 
``\onefinite'' case) and a second that works for any arbitrary configuration of 
source and observer (the ``\twofinite'' case). Furthermore, we demonstrate that 
the complete solution can be expressed using no more than 3 roots in the 
{\twofinite} case and no more than 2 roots in the {\onefinite} case. We 
apply an Euler rotation and make use of the particulars of the resulting 
geometry to simplify the analysis, which also serves to minimize the associated 
computational expenses. 

\section{Euler Rotation} \label{sec:EulerRotation}

In both cases, it is useful to place either the source or observer at polar angle 
$\sfrac{\pi}{2}$ (i.e. above the north pole) for two reasons: first, it reduces the 
subsequent analysis to two dimensions and second, it allows a simplification which 
will be used in Sections \ref{sec:OneFinite} and \ref{sec:TwoFinite}. The simplest 
way to accomplish this is to start in Cartesian coordinates with the origin at the 
center of the sphere and then apply two sequential rotations. The first rotation uses
the angles

\begin{equation}
	\alpha = \frac{\pi}{2} - \atantwo{y_{\src}, x_{\src}} \tab \text{and} \tab 
	\beta = \frac{\pi}{2} - \asin{\frac{z_{\src}}{R_{\src}}}
\end{equation}
to define new coordinates in a rotated frame (denoted by the prime, $'$),
\begin{subequations} \label{eqn:EulerRotation1}
	\begin{equation}
		x_k' = z_k \psin{\beta} - x_k \pcos{\beta} \psin{\alpha} - y_k \pcos{\beta} \pcos{\alpha}, 
	\end{equation}
	\begin{equation}
		y_k' = x_k \pcos{\alpha} + y_k \psin{\alpha}, \text{ and}
	\end{equation}
	\begin{equation}
		z_k' = x_k \psin{\beta} \psin{\alpha} + y_k \pcos{\alpha} \psin{\beta} + z_k \pcos{\beta}. 
	\end{equation}
\end{subequations}
Here $\alpha$ is the azimuthal angle of the source (measured from the positive 
$x$-axis), and $\beta$ is the inclination angle (measured from the positive 
$z$-axis). The second rotation is not strictly necessary; the critical portion 
is the placement of source or observer at $\sfrac{\pi}{2}$, accomplished by the 
first rotation. However, this second rotation further reduces the problem 
because it enforces $-\sfrac{\pi}{2} < \theta_{\obs} < \sfrac{\pi}{2}$ (in the 
rotated frame), allowing us to ignore some of the roots in Equations 
\ref{eqn:OneFinite_Coeff} and \ref{eqn:TwoFinite_coeff}. This rotation takes
\begin{equation}
	\gamma = \atantwo{y_{\src}', x_{\src}'}  
\end{equation}
to establish coordinates in the twice-rotated, $''$, frame to be 
\begin{subequations} \label{eqn:EulerRotation2}
	\begin{equation}
		x_k'' = z_k' \psin{\gamma} - y_k' \pcos{\gamma}, 
	\end{equation}
	\begin{equation}
		y_k'' = y_k' \psin{\gamma} - z_k' \pcos{\gamma}, \text{ and}
	\end{equation}
	\begin{equation}
		z_k'' = x_k'.
	\end{equation}
\end{subequations}
With $k \in \{\src, \obs\}$ (i.e. for $x_{\src}, ~ x_{\obs}$, etc) and where 
$\atantwo{y, x}$ defines the angle from the positive $x$-axis to $(x, y)$. 
At the end, once the specular point has been calculated in this new 
coordinate system, we will apply these two rotation in reverse for $k \in 
\{\src, \obs, \spec\}$ to recover the specular point in the original system. 
The rotated system $(x_k'', y_k'', z_k'')$ will be used for the remainder of 
this paper.

\renewcommand*{\arraystretch}{2.25}
In the {\onefinite} case, the rotation should be performed to place the 
coordinate of \emph{finite} radius at $\sfrac{\pi}{2}$. This allows full 
use to be made of the approximation $R_{\obs} \gg R_{\sph}$ or $R_{\src} 
\gg R_{\sph}$. In the {\twofinite} case the rotation can place either 
coordinate at $\sfrac{\pi}{2}$, but we will place the source at 
$\sfrac{\pi}{2}$ for notational consistency.  

{\color{onefinite} \section{One-Finite Case} \label{sec:OneFinite}}
If either the source or the observer can be considered infinitely 
distant, then after applying the Euler rotation in Equation 
\ref{eqn:EulerRotation1} to place the finitely distant of the two 
at $\sfrac{\pi}{2}$ -- as Figure \ref{fig:SphericalSpecular} shows 
-- $\theta_1$ and $\theta_2$ can be defined such that
\begin{subequations} \label{eqn:OneFinite_thetas}
	\begin{equation}
		\theta_1 = \atan{\frac{R_{\obs}  \psin{\theta_{\obs}} - R_{\sph} \psin{\theta_{\spec}}}{R_{\obs} \pcos{\theta_{\obs}} - R_{\sph} \pcos{\theta_{\spec}}}} \tab \text{and}
	\end{equation}
	\begin{equation} \label{eqn:theta2}
		\theta_2 = \atan{\frac{R_{\src}  \psin{\theta_{\src}} - R_{\sph} \psin{\theta_{\spec}}}{R_{\src} \pcos{\theta_{\src}} - R_{\sph} \pcos{\theta_{\spec}}}}, 
	\end{equation}
\end{subequations}
where $\theta_e$ and $\theta_i$ are then given by
\begin{equation*}
	\theta_e = \frac{\pi}{2} - \left( \theta_{\spec} - \theta_1 \right) \tab \text{and} \tab 
	\theta_i = \frac{\pi}{2} - \left( \theta_2 - \theta_{\spec} \right).
\end{equation*}

\subsection{Formulation}
Since incidence angle ($\theta_i$) must always be equal to emission angle 
($\theta_e$), then in the limit where $R_{\obs} \gg R_{\sph}$ we get
\begin{equation} \label{eqn:OneFinite_theta2}
	\theta_2 = 2 \theta_{\spec} - \theta_{\obs}.
\end{equation}
With the introduction of a constant, 
\begin{equation}
	c \equiv \frac{R_{\sph}}{R_{\src}}, 
\end{equation}
and substituting in $\theta_2$ from Equation
\ref{eqn:theta2}, Equation \ref{eqn:OneFinite_theta2} becomes
\begin{equation} \label{eqn:OneFinite_initial}
	\frac{\psin{\theta_{\src}} - c \psin{\theta_{\spec}}}{\pcos{\theta_{\src}} - c \pcos{\theta_{\spec}}} = \ptan{2 \theta_{\spec} - \theta_{\obs}}.
\end{equation}
Making use of the identities
\begin{equation*}
	\ptan{x - y} = \frac{\psin{2x} - \psin{2y}}{\pcos{2x} + \pcos{2y}} \tab \text{and} \tab
	\sin x \cos 2y \pm \sin 2y \cos x = \psin{2 y \pm x}, 
\end{equation*}
and the fact that the Euler rotation in Equation \ref{eqn:EulerRotation1} results in $\theta_{\src} = \sfrac{\pi}{2}$, we get
\begin{equation} \label{eqn:SpecMillersEquation}
	\pcos{4 \theta_{\spec}} + c \psin{3 \theta_{\spec}} + \pcos{2 \theta_{\obs}} = c \psin{\theta_{\spec} - 2 \theta_{\obs}}.
\end{equation}

\subsection{Solution}
Equation \ref{eqn:SpecMillersEquation} has sixteen unique, non-trivial 
solutions, courtesy of Wolfram Mathematica \citep{Mathematica}. We neglect 
eight of these because they represent special cases that are not 
physically meaningful for our purposes. The other eight are presented in 
Equation \ref{eqn:OneFinite_thetap} using the coefficients we define in 
Equation \ref{eqn:OneFinite_Coeff}.
\begin{subequations} \label{eqn:OneFinite_Coeff}
	\begin{align}
		D_0 \equiv ~ & 1152 \left( c^2 - 4 \right) \big( c^2 + c^2 \pcos{2 \theta_{\obs}} - %\\
			      846 \pcos{2 \theta_{\obs}} - 1 \big) +  c^2 \psin{2 \theta_{\obs}}^2 \\[4mm]
		D_1 \equiv ~ & \bigg( \minus 4 \big( 160 - 128 c^2 + 4 c^4 + 96 \pcos{2 \theta_{\obs}} - %\\
			      96 c^2 \pcos{2 \theta_{\obs}} \big)^3 + \left( \minus 16 \left( c^2 - 4\right)^3 - D_0 \right)^2 \bigg)^{\sfrac{1}{2}} \\[4mm]
		D_2 \equiv ~ & \sqrt[3]{16 \left( c^2 - 4\right)^3 + D_0 - D_1} \\[4mm]
		D_3 \equiv ~ & \frac{1}{3 \sqrt[3]{4} D_2} \Big( 40 - 32 c^2 + c^4 + 24 \pcos{2 \theta_{\obs}} - %\\
			      24 c^2 \pcos{2 \theta_{\obs}} \Big) + \frac{D_2}{24 \sqrt[3]{2}} \\[4mm]
		D_4 \equiv ~ & \sqrt{\frac{4 - c^2}{6} + D_3}
	\end{align}
\end{subequations}

\begin{equation} \label{eqn:OneFinite_thetap}
	\theta_{\spec} = \pm \left\{ \begin{array}{c}
		\displaystyle \acos{- \frac{D_4}{2} - \sqrt{\minus \frac{4 - c^2}{3} - D_3 - \frac{c \psin{2 \theta_{\obs}}}{8 D_4}}} \\
		\displaystyle \acos{- \frac{D_4}{2} + \sqrt{\minus \frac{4 - c^2}{3} - D_3 - \frac{c \psin{2 \theta_{\obs}}}{8 D_4}}} \\
		\displaystyle \acos{+ \frac{D_4}{2} - \sqrt{\minus \frac{4 - c^2}{3} - D_3 + \frac{c \psin{2 \theta_{\obs}}}{8 D_4}}} \\
		\displaystyle \acos{+ \frac{D_4}{2} + \sqrt{\minus \frac{4 - c^2}{3} - D_3 + \frac{c \psin{2 \theta_{\obs}}}{8 D_4}}} \\
	\end{array} \right.
\end{equation}

\begin{figure*}[b]
	\includegraphics[width=\textwidth, trim={3mm, 2mm, 3mm, 2mm}]{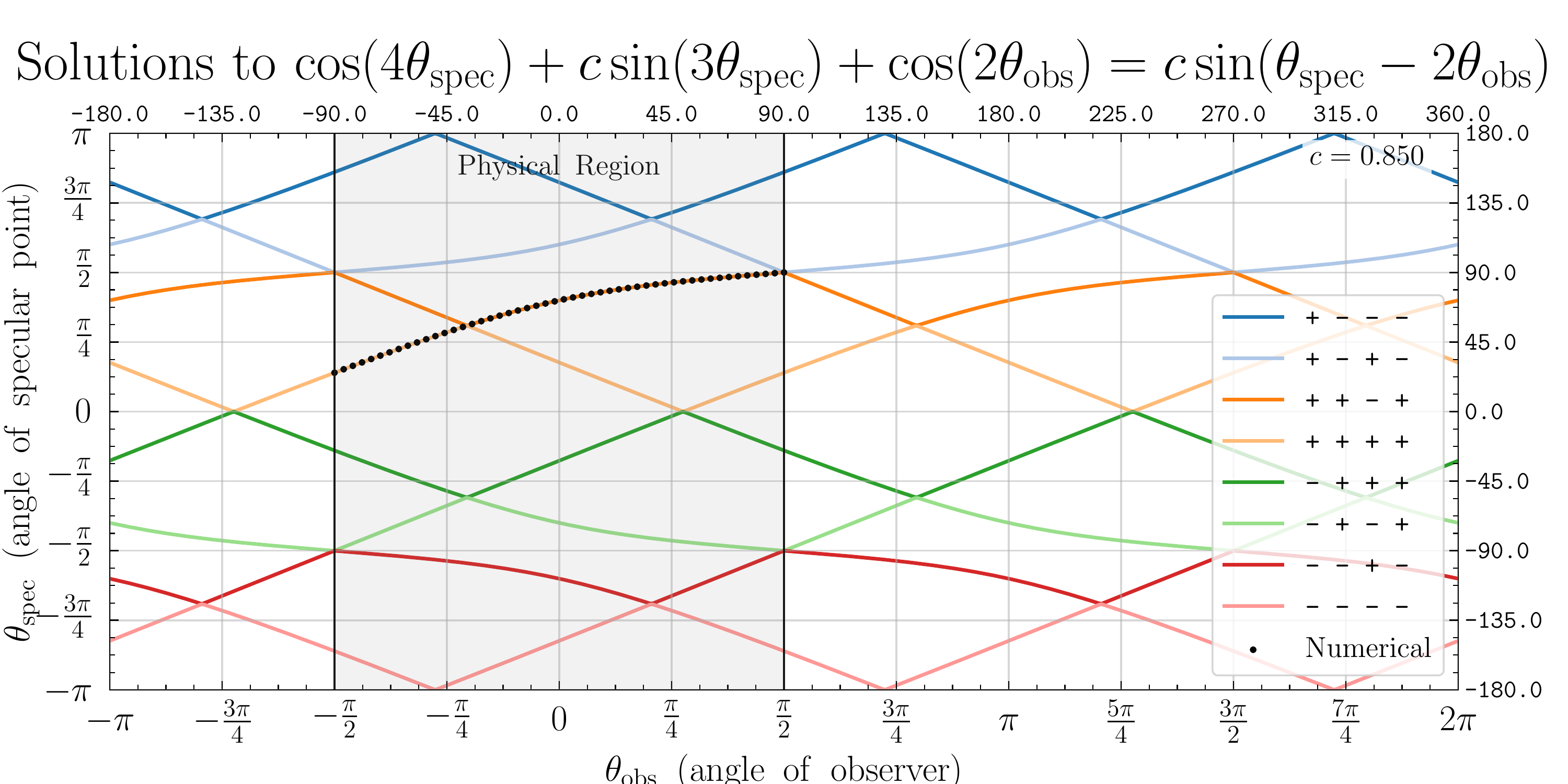}
	\caption{Plots of the solutions to Equation \ref{eqn:SpecMillersEquation} where 
			 $R_{\sph} / R_{\src}$ ($c$) is $0.85$. Note that the numerical method is only applied to 
			 $[\minus \sfrac{\pi}{2}, \sfrac{\pi}{2}]$ because the Euler 
			 rotation in Section \ref{sec:EulerRotation} enforces $\theta_{\obs}$
			 to be between $\minus \sfrac{\pi}{2}$ and $\sfrac{\pi}{2}$.
			 $\theta_{\obs}$ is the angle of the observer, $\theta_{\spec}$ is 
			 the angle of the specular point. Both are in the rotated reference
			 frame and measured from the positive $x$-axis. An animated version
			 is available which shows the solutions for various values of $c$.}
	\label{fig:OneFinite_Specular-0.85}
\end{figure*}

The only differences between the solutions are the different signs 
associated with four terms. The possible combinations are $\pm - - -$, 
$\pm - + -$, $\pm + - +$, and $\pm + + +$. These solutions are plotted 
in Figure \ref{fig:OneFinite_Specular-0.85} for a source at 3000 km 
from the center of a sphere of radius 2575 km (i.e. Titan) and observer 
at much greater distance. After comparing these solutions against the 
numerical method, as in Figure \ref{fig:OneFinite_Specular-0.85}, it is 
clear the branch that produces the physical specular point (on the 
exterior of the sphere) is
\begin{equation}
	\theta_{\spec} = \left\{ \begin{array}{lrcl}
		\displaystyle \acos{\frac{D_4}{2} - \sqrt{\minus \frac{4 - c^2}{3} - D_3 + \frac{c \psin{2 \theta_{\obs}}}{8 D_4}}} &
			\tab \theta_{\spec} & < & \displaystyle \frac{\pi ( \acot{c} - \theta_{\obs})}{\pi + 2 \acot{c}} \\
		\displaystyle \acos{\frac{D_4}{2} + \sqrt{\minus \frac{4 - c^2}{3} - D_3 + \frac{c \psin{2 \theta_{\obs}}}{8 D_4}}} &
			\tab \theta_{\spec} & \geq & \displaystyle \frac{\pi ( \acot{c} - \theta_{\obs})}{\pi + 2 \acot{c}} \\
	\end{array} \right.
\end{equation}
% \subsection{Observer at finite distance}
% 
% By altering the Euler rotation in \ref{eqn:EulerRotation1} to place the observer at 
% $\theta_{\obs} = \sfrac{\pi}{2}$ instead of the source, this same approach can be 
% used in the limit where $R_{\src} \gg R_{\sph}$ and $R_{\obs} \not\gg R_{\sph}$. In 
% this case Equation \ref{eqn:SpecMillersEquation} becomes
% \begin{equation} \label{eqn:SpecMillersEquation_DistantSource}
% 	\pcos{4 \theta_{\spec}} + c \psin{3 \theta_{\spec}} + \pcos{2 \theta_{\src}} = c \psin{\theta_{\spec} - 2 \theta_{\src}}	
% \end{equation} 
% and the solutions thereof will similarly have $\theta_{\obs}$ replaced with $\theta_{\src}$.

\pagebreak
{\color{twofinite} \section{Both-Finite Case} \label{sec:TwoFinite}}
\subsection{Formulation}
If $R_{\obs} \not\gg R_{\sph}$ and $R_{\src} \not\gg R_{\sph}$ then the solution 
becomes even more complicated. In this case, Equation \ref{eqn:OneFinite_theta2} 
becomes
\begin{equation}
	\theta_2 + \theta_1 = 2 \theta_{\spec}, 
\end{equation}
and Equation \ref{eqn:OneFinite_thetas} becomes
\begin{subequations}
	\begin{equation}
		\theta_1 = \atan{\frac{R_{\obs} \psin{\theta_{\src}} - R_{\sph} \psin{\theta_{\spec}}}{R_{\obs} \pcos{\theta_{\src}} - R_{\sph} \pcos{\theta_{\spec}}}} \tab \text{and}
	\end{equation}
	\begin{equation}
		\theta_2 = \atan{\frac{R_{\src} \psin{\theta_{\obs}} - R_{\sph} \psin{\theta_{\spec}}}{R_{\src} \pcos{\theta_{\obs}} - R_{\obs} \pcos{\theta_{\spec}}}}, \tab
	\end{equation}
\end{subequations}
which yields
\begin{equation*}
	\atan{\frac{R_{\obs} \psin{\theta_{\src}} - R_{\sph} \psin{\theta_{\spec}}}{R_{\obs} \pcos{\theta_{\src}} - R_{\sph} \pcos{\theta_{\spec}}}}
	 + \atan{\frac{R_{\src} \psin{\theta_{\obs}} - R_{\sph} \psin{\theta_{\spec}}}{R_{\src} \pcos{\theta_{\obs}} - R_{\obs} \pcos{\theta_{\spec}}}} = 2 \theta_{\spec} ~. 
\end{equation*}
Again making use of $\theta_{\src} = \sfrac{\pi}{2}$ and defining a new constant, 
\begin{equation}
	b \equiv \frac{R_{\sph}}{R_{\obs}}, 
\end{equation} 
the {\twofinite} equivalent of Equation \ref{eqn:SpecMillersEquation} is
\begin{equation} \label{eqn:MillersEquation}
	\frac{c \psin{\theta_{\obs} + \theta_{\spec}} - c b \psin{2 \theta_{\spec}} + \cos \theta_{\obs} - b \cos \theta_{\spec}}
	     {c \pcos{\theta_{\obs} + \theta_{\spec}} - c b \pcos{2 \theta_{\spec}} + \sin \theta_{\obs} - b \sin \theta_{\spec}} = \ptan{2 \theta_{\spec}} ~ .
\end{equation}
With $b = 0$ (and with some manipulation) we can recover Equation 
\ref{eqn:SpecMillersEquation}. Doing so is nontrivial and will be neglected for 
brevity, but it is clearly demonstrated by the agreement between the two methods 
when $b \approx 0$ as shown in section \ref{sec:precision}.

\subsection{Solution}
Similar to Equation \ref{eqn:SpecMillersEquation}, Equation \ref{eqn:MillersEquation} 
produces sixteen unique, non-trivial solutions -- only eight of which are physically 
meaningful (the others represent special cases). These are presented in Equation 
\ref{eqn:TwoFinite_thetap} using the coefficients defined in Equation 
\ref{eqn:TwoFinite_coeff}
\begin{subequations} \label{eqn:TwoFinite_coeff}
	\begin{align}
		E_0 \equiv ~& b^2 + c^2 + 2 b c \sin \theta_{\obs} - 4 \\[3mm]
		E_1 \equiv ~& 24 \pcos{\theta_{\obs}}^2 \left(b^2 - b c \sin \theta_{\obs} - 2 c^2 + 2 \right) + E_0^2 \\[3mm]
		E_2 \equiv ~& 72 E_0 \cos^2 \theta_{\obs} \left(b^2 - b c \sin \theta_{\obs} + 4 c^2 - 4 \right) +
		      432 \cos^2 \theta_{\obs} \Big( (b - c \sin \theta_{\obs} )^2 -
		      b^2 (c^2 - 1) \cos^2 \theta_{\obs} \Big) + 2 E_0^3 \\[3mm]
		E_3 \equiv ~& \left( \sqrt{E_2^2 - 4 E_1^3} + E_2 \right)^{\sfrac{1}{3}} \\[3mm]
		E_4 \equiv ~& \frac{\sqrt[3]{4}\, E_3^2 + \sqrt[3]{16}\, E_1}{24 E_3} \\[3mm]
		E_5 \equiv ~& \sqrt{\frac{b^2 \pcos{\theta_{\obs}}^2}{4} + \frac{E_3}{12 \sqrt[3]{2}}+\frac{E_1}{6 \sqrt[3]{4}\ E_3} - \frac{E_0}{6}} \\[3mm]
		E_6 \equiv ~& \frac{(b \cos \theta_{\obs})^3 - b E_0 \cos \theta_{\obs} - 4 \cos \theta_{\obs} \left( b - c \sin \theta_{\obs} \right)}{16 E_5}
	\end{align}
\end{subequations}

\begin{equation} \label{eqn:TwoFinite_thetap}
	\theta_{\spec} = \pm \left\{ \begin{array}{l}
		\displaystyle \acos{ \frac{b \cos \theta_{\obs} - 2 E_5}{4} - \frac{1}{2} \sqrt{ \frac{3 (b \cos \theta_{\obs})^2 - 2 E_0}{12} - E_4 - E_6}} \\
		\displaystyle \acos{ \frac{b \cos \theta_{\obs} - 2 E_5}{4} + \frac{1}{2} \sqrt{ \frac{3 (b \cos \theta_{\obs})^2 - 2 E_0}{12} - E_4 - E_6}} \\
		\displaystyle \acos{ \frac{b \cos \theta_{\obs} + 2 E_5}{4} - \frac{1}{2} \sqrt{ \frac{3 (b \cos \theta_{\obs})^2 - 2 E_0}{12} - E_4 + E_6}} \\
		\displaystyle \acos{ \frac{b \cos \theta_{\obs} + 2 E_5}{4} + \frac{1}{2} \sqrt{ \frac{3 (b \cos \theta_{\obs})^2 - 2 E_0}{12} - E_4 + E_6}}
	\end{array} \right.
\end{equation}

\begin{figure}[t]
	\includegraphics[width=\textwidth, trim={3mm, 2mm, 3mm, 2mm}]{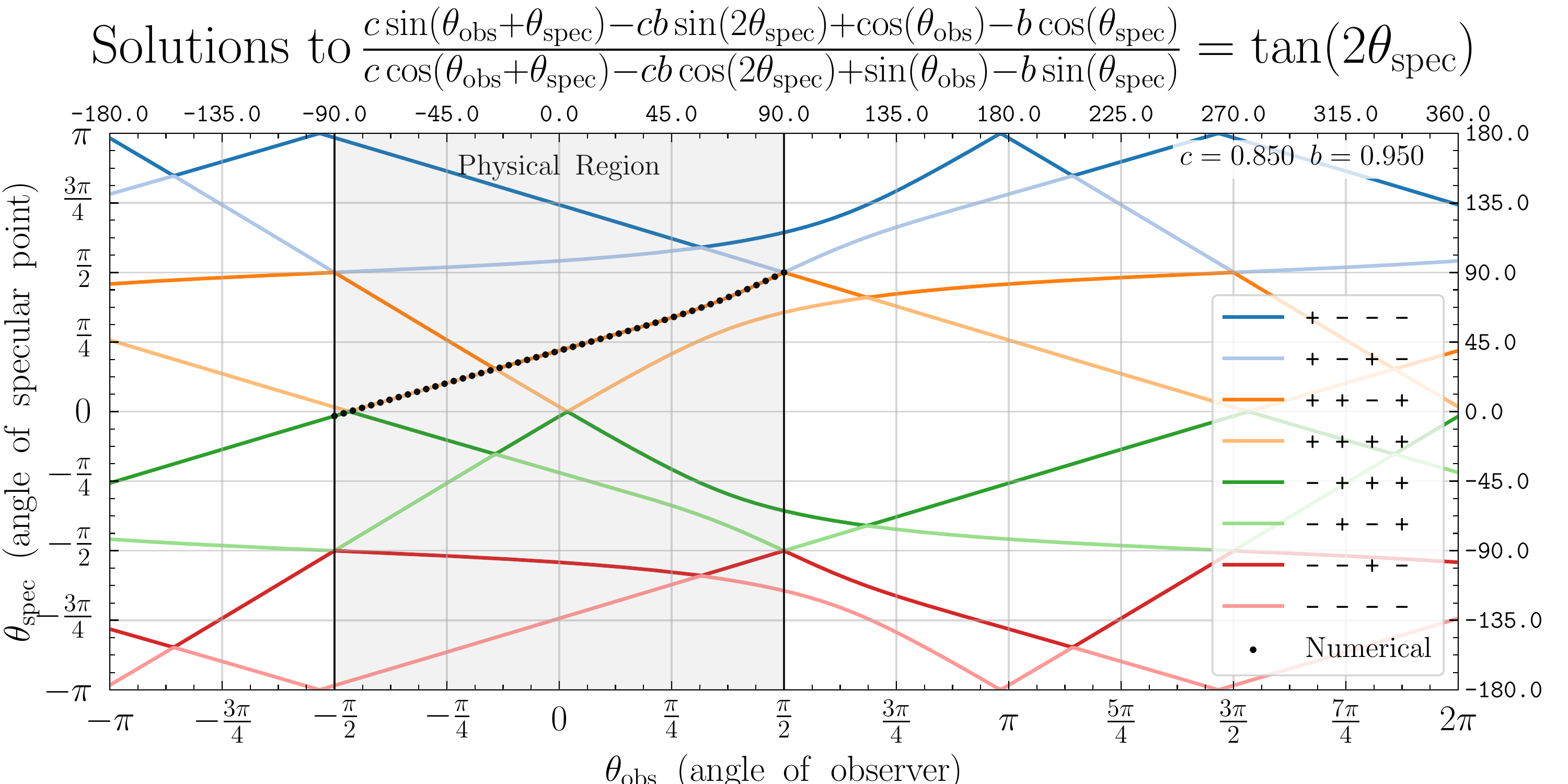}
	\caption{Solutions to Equation \ref{eqn:TwoFinite_thetap} for $R_{\sph} / R_{\src}$ ($c$) of $0.85$
			 and $R_{\sph} / R_{\obs}$ ($b$) of $0.95$. Note that the numerical method is only applied to 
			 $[\minus \sfrac{\pi}{2}, \sfrac{\pi}{2}]$ because the Euler 
			 rotation in Section \ref{sec:EulerRotation} enforces $\theta_{\obs}$
			 to be between $\minus \sfrac{\pi}{2}$ and $\sfrac{\pi}{2}$.
			 $\theta_{\obs}$ is the angle of the observer, $\theta_{\spec}$ is 
			 the angle of the specular point. Both are in the rotated reference
			 frame and measured from the positive $x$-axis. An animated version
			 is available which shows the solutions for various values of $b$ and $c$.} 
	\label{fig:SpecularPlot-0.95-0.85}
\end{figure}

As in Equation \ref{eqn:OneFinite_thetap}, the only differences between the solutions 
are the signs on four terms. The combinations of these signs are $\pm - - -$, 
$\pm - + -$, $\pm + - +$, and $\pm + + +$ (the same as in \ref{eqn:OneFinite_thetap}). 
They are plotted in Figure \ref{fig:SpecularPlot-0.95-0.85}. Comparison with the 
numerical method indicates the branch that produces the physical specular point (on 
the exterior of the sphere) is

\begin{equation*}
	E_7 \equiv \frac{3 (b \cos \theta_{\obs})^2 - 2 E_0}{12} - E_4 + E_6
\end{equation*}
\begin{equation} \label{eqn:TwoFinite_soln}
	\theta_{\spec} = \left\{ \begin{array}{cccc}
		\displaystyle \minus \acos{ \frac{b \cos \theta_{\obs} + 2 E_5}{4} + \frac{1}{2} \sqrt{E_7}} &
			\tab \minus \dfrac{\pi}{2}  \leq  \theta_{\obs} < L_1 \\
		\displaystyle \acos{ \frac{b \cos \theta_{\obs} + 2 E_5}{4} + \frac{1}{2} \sqrt{E_7}} &
			\tab L_1 \leq \theta_{\obs} \leq L_2 &  \text{ and } & \displaystyle \frac{\partial \, E_7}{\partial \, \theta_{\obs}} \leq 0  \\
		\displaystyle \acos{ \frac{b \cos \theta_{\obs} + 2 E_5}{4} - \frac{1}{2} \sqrt{E_7}} & 
			\tab L_2 < \theta_{\obs} \leq \dfrac{\pi}{2} & \text{ or } & \displaystyle \frac{\partial \, E_7}{\partial \, \theta_{\obs}} > 0 \\  
	\end{array} \right.
\end{equation}

where the limits $L_1$ and $L_2$ are given by
\begin{subequations}
	\begin{equation}
		L_1 = \atan{\frac{c + b \sqrt{1 - b^2 + c^2}}{b^2 - c^2}}
	\end{equation}
	\begin{equation}
		L_2 = \atan{\frac{c - b \sqrt{1 - b^2 + c^2}}{b^2 - c^2}}
	\end{equation}
\end{subequations}
It should be noted that $L_1$ may occur at less than $\minus \sfrac{\pi}{2}$, and 
in such cases only the second two parts in Equation \ref{eqn:TwoFinite_soln} are 
necessary. Additionally, while there is an exact solution for the partial 
derivative of $E_7$, in practical computation it is robust to use the approximation
\begin{equation*}
	\frac{\partial \, E_7}{\partial \, \theta_{\obs}} \approx E_7\big\rvert_{\theta_{\obs}} - 
	E_7\big\rvert_{\theta_{\obs} - \zeta}
\end{equation*}
where $\zeta$ is directly related to the branch-deduction precision (assuming it is 
larger than the computation precision), e.g. $\zeta = 10^{\minus 9}$ will result in 
the correct branch being chosen within $10^{\minus 9}$. Our {\twofinite} 
implementation in the following section makes use of this approach. 

\begin{figure}[b]
	\centering
	\includegraphics[width=\textwidth]{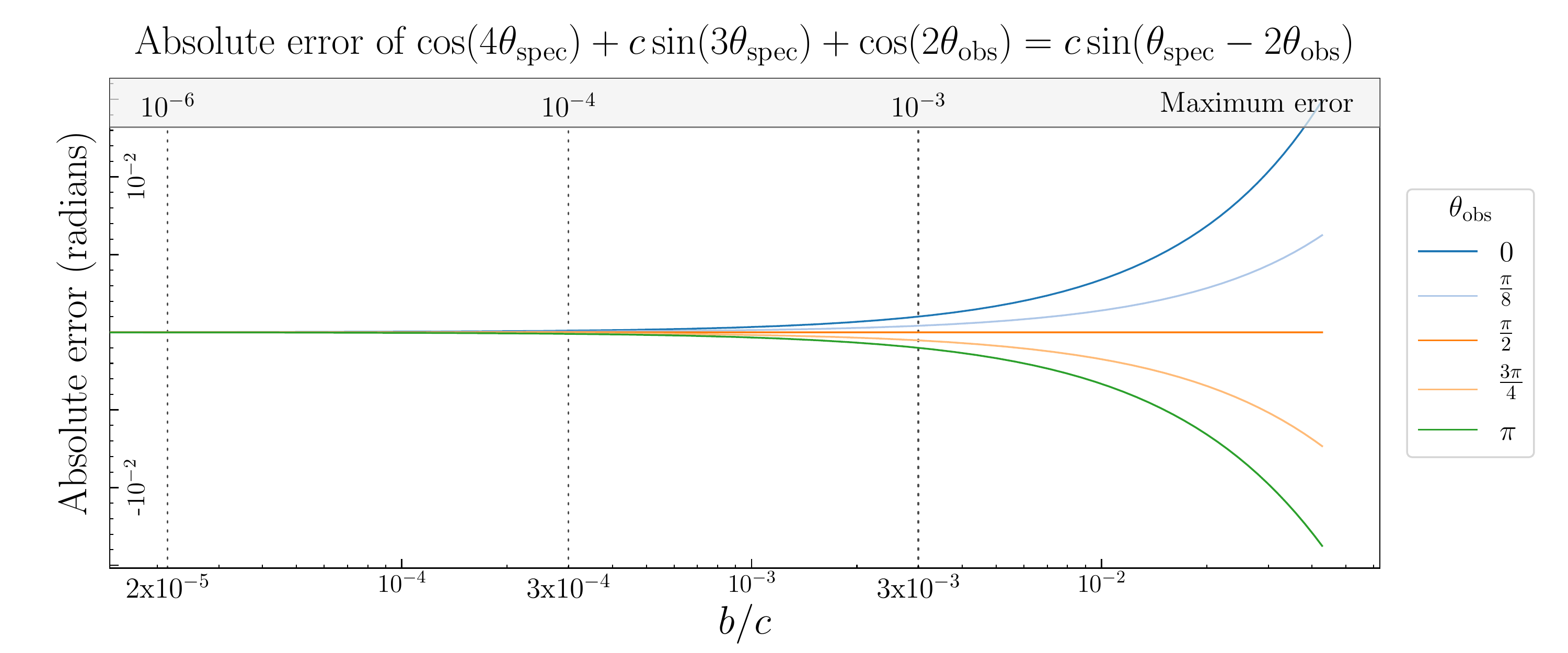}
	\caption{The agreement between the {\onefinite} and {\twofinite} solutions
	         as a function of the ratio between $b$ and $c$ with the maximum
	         error across all $\theta_{\obs}$ for a given $b / c$. Note that
	         even though $\theta_{\obs}$ is confined to $[\minus \sfrac{\pi}{2}, 
	         \sfrac{\pi}{2}]$, we are showing the error for $\theta_{\obs}$ 
	         between $0$ and $\pi$. We do so to demonstrate that for 
	         $-\sfrac{\pi}{2} > \theta_{\obs}$ or $\theta_{\obs} < \sfrac{\pi}{2}$ 
	         (in the reference frame before the second Euler rotation) the 
	         persistent error is 
	         symmetrically reversed.}
	\label{fig:ErrorPlot}
\end{figure}

\section{Computation} \label{sec:Computation}

Implementations of these solutions in both Python and C\lstinline{++}{} can be 
found in \href{https://github.com/WMiller256/Alhazen-Ptolemy}{this GitHub 
Repository}\footnote{https://github.com/WMiller256/Alhazen-Ptolemy}. Also in 
the repository are Wolfram Language Package files for easily reading the solutions 
into Mathematica. The C\lstinline{++}{} implementations were used for performance 
benchmarking. The solution for the {\onefinite} case performs significantly better 
than that of the {\twofinite} case, but in many applications the difference will 
be negligible because both routines are relatively minimal -- especially 
compared to numerical methods.

\subsection{Performance}

Excluding branch deductions (and after compiling with \lstinline{gcc}{} option 
\lstinline{-O3}{}), our C\lstinline{++}{} implementation of the 
{\onefinite} solution requires 76 floating point operations, 6 trigonometric and 5 
\lstinline{sqrt}{} calls with one expensive exponentiation operation to find a 
complex cube root. The corresponding {\twofinite} code requires 131 
floating point operations, 5 trigonometric and 5 \lstinline{sqrt}{} calls, and one 
expensive cube root. 

Benchmarked on an hexacore Intel i7-8750H the {\onefinite} solution averaged $23$ 
ns per iteration and the {\twofinite} solution averaged $121$ ns per iteration for 
$10^9$ random combinations of $b$, $c$, and $\theta_{\obs}$. The code used for 
benchmarking is included in the linked GitHub repository. In cases where 
$L_1 \leq \theta_{\obs} \leq L_2$ the {\onefinite} implementation has a 
more significant speed advantage because the {\twofinite} implementation's 
branch deductions are most expensive in this region.

\subsection{Precision} \label{sec:precision}

The agreement between the {\onefinite} solution and the {\twofinite} solution
is shown in Figure \ref{fig:ErrorPlot}. When using the 64-bit \lstinline{double}{} type, 
the machine precision for either calculation is roughly $10^{\minus 7}$
due to the trigonometric functions. The use of the \lstinline{long double}{} type (which 
is implementation dependent but usually corresponds to between 80-bit and 128-bit 
storage) can be used to improve this precision to $10^{\minus 15}$. It should
be noted that any numerical methods will have the same or similar precision limitations.
Figure \ref{fig:ErrorPlot} does not make use of the extra precision and therefore 
the point at which the average error exceeds $10^{\minus 6}$ can be considered the 
point at which the disagreement between the two methods exceeds the machine 
precision. This occurs at around $\sfrac{b}{c} = \sci{2}{\minus 5}$, i.e. when either 
$R_{\obs}$ or $R_{\src}$ is 50,000 times larger than the other. At a ratio of 
500 the accuracy drops to $\pm \sci{1.72}{\minus 2}$ degrees, but only for the nearly 
worst-case-scenario of $c = 0.95$. As $c$ decreases the accuracy of $\theta_{\spec}$ 
at the same $\sfrac{b}{c}$ improves, e.g. for $c=0.1$ the accuracy is 
$\pm \sci{1.72}{\minus 6}$ degrees at $\sfrac{b}{c} = 500$. 

\section{Conclusion}

\subsection{SRTC++}

Our method is of particular use in simulating radiative transfer on Titan, i.e. in
the \lstinline{SRTC++}{} model \citep{barnes2018spherical}. Specifically, calculating 
specular reflections due to atmospheric scattering allows us to compensate for the 
adjacency effect -- when atmospheric scattering redirects light reflected off bright 
material adjacent to spectrally dark regions (e.g. lakes) and causes them to appear 
brighter than they should. This effect has been well documented on Earth 
\citep{odell1975effect, tanre1981influence, minomura2001atmospheric, 
sterckx2011detection, kiselev2015sensor} and \citet{karkoschka2016eight} found it 
necessary to compensate for it in their reflectivity analysis of \textit{Huygens} data.

Realistically simulating Titan's atmosphere and surface requires compensation for
the adjacency effect, and many high fidelity simulations. In such
simulations, the specular point must be calculated for every scattering event (and 
there are usually billions). The effect of our method on the computation speed of 
these simulations is negligible (e.g. about 1 minute over 4 days of execution 
time). The same cannot be said of existing methods. 

\subsection{Computer rendering}

Another notable application is in rendering computer graphics. Computation speed is
critical here as well, and often multiple light sources or reflections must be 
accounted for -- each with a separate specular point. Sources with non-negligible 
spatial extent are also common, requiring the calculation of multiple specular points
during each refresh. 
We have formulated this paper to emphasize spherical surfaces, but in fact the 
analysis in the rotated reference frame can be applied to any reflective surfaces
of revolution, provided the intersection between the surface and the plane formed
by the source, observer, and specular point forms a circle.   

\subsection{Extension to elliptical geometry}

Extending our approach to elliptical geometry would drastically broaden its usefulness 
by providing an analytical, branch-deducing method for calculating specular points on 
any closed conic section. As mentioned previously, this extension would enable our 
solution to be  used in GPS communications. But it would also expand the usefulness 
for computer rendering because most planar slices of regular surfaces of revolution 
produce one or several non-circular conic sections. We leave the elliptical case to 
future work.

\subsection{In Summary}

We have presented a formulation of the Alhazen-Ptolemy problem in which the physical 
specular point is predeterminable, we find this formulation to be superior to previous 
approaches because it does not rely on numerical methods for branch deduction. 
Furthermore, we have shown that the solution can be directly written as a piecewise 
function with only 3 branches in the {\twofinite} case and only 2 branches in the 
{\onefinite} case. Employing an Euler rotation, normalizing the radius of the sphere 
to unity, and translating the rotated coordinate system to center on the specular 
point are critical in this formulation and in enabling robust and efficient 
computational implementation. Our method is useful for studying the ethane 
composition of Titan's lakes with \lstinline{SRTC++}{}, rendering 3D computer 
graphics, and lays the groundwork for extension to elliptical geometry.

\bibliography{AlhazenPtolemy}
\bibliographystyle{aasjournal}
 
\end{document}